\documentclass[]{ceurart}

\usepackage{listings}
\usepackage{booktabs}

\usepackage{acro}
\DeclareAcronym{RDF}{
	short = RDF,
	long  = Resource Description Framework,
}
\DeclareAcronym{CI}{
	short = CI,
	long  = Continuous Integration,
}
\DeclareAcronym{IRI}{
	short = IRI,
	long  = Internationalized Resource Identifier,
}
\DeclareAcronym{SOP}{
	short = SOP,
	long  = Semantic Operation Pipeline,
}
\DeclareAcronym{HDT}{
	short = HDT,
	long  = {Header information, a Dictionary, and the actual Triples structure},
}

\usepackage[dvipsnames,svgnames]{xcolor} 
\usepackage[normalem]{ulem} 

\makeatletter
\font\uwavefont=lasyb10 scaled 700
\def\spelling{\bgroup\markoverwith{\lower3.5\p@\hbox{\uwavefont\textcolor{Red}{\char58}}}\ULon}
\def\grammar{\bgroup\markoverwith{\lower3.5\p@\hbox{\uwavefont\textcolor{LimeGreen}{\char58}}}\ULon}
\def\phrasing{\bgroup\markoverwith{\lower3.5\p@\hbox{\uwavefont\textcolor{RoyalBlue}{\char58}}}\ULon}

\newcommand\remove{\bgroup\markoverwith{\textcolor{red}{\rule[0.5ex]{2pt}{0.4pt}}}\ULon}
\newcommand\insertion{\bgroup\markoverwith{\textcolor{Green}{\rule[-0.5ex]{2pt}{0.6pt}}}\ULon}
\makeatother

\def\arxivversion{}
\newif\ifarxiv
\ifdefined\arxivversion\arxivtrue\else\arxivfalse\fi

\begin{document}

\copyrightyear{2026}
\copyrightclause{Copyright for this paper by its authors.
	Use permitted under Creative Commons License Attribution 4.0 International (CC BY 4.0).}

\ifarxiv
\conference{}
\else
\conference{DMKG 2026: 2nd International Workshop on Data Management for Knowledge Graphs,
	October 25--26, 2026, co-located with ISWC 2026, Bari, Italy}
\fi

\title{Strengthening LargeRDFBench for Interoperable Federated SPARQL Evaluation}

\author[1]{Bryan-Elliott Tam}[%
	orcid=0000-0003-3467-9755,
	email=bryanelliott.tam@ugent.be,
]
\cormark[1]
\author[2]{Muhammad Saleem}[%
    orcid=0000-0001-9648-5417,
	email=saleem.muhammd@gmail.com,
]
\author[1]{Ruben Taelman}[%
	orcid=0000-0001-5118-256X,
	email=ruben.taelman@ugent.be,
]
\address[1]{Department of Electronics and Information Systems, Ghent University -- imec}
\address[2]{AKSW, University Leipzig, Germany}
\cortext[1]{Corresponding author.}

\begin{abstract}
	LargeRDFBench is one of the most comprehensive benchmarks for evaluating federated SPARQL
	query engines, combining a large collection of real, interlinked datasets with a rich query
	suite that has made it a reference point for the community.
	Evaluations of federated engines are published by comparing engine results against the benchmark's expected results, so it is important that the results of the benchmark are themselves reproducible.
	Moreover, several of its data dumps violate the RDF specifications, so only engines that parse RDF leniently can host them, and its expected results are distributed in an ad hoc format.
	We systematically identify and categorize these data-quality issues and repair them,
	producing standards-conformant serializations of every affected dataset.
	Furthermore, we re-encode the benchmark's expected results in the W3C SPARQL 1.1 Query Results JSON Format and correct their discrepancies.
	We contribute a standards-compliant edition of LargeRDFBench, produced by a
	reproducible cleaning pipeline, together with its expected query results in a standard,
	machine-verifiable format.
	Every dataset now parses under strict, specification-compliant RDF parsers, and the expected
	results are machine-verifiable through a standard format, extending the benchmark's reach to
	the full range of conformant engines while staying faithful to the original data.
	Reproducing the expected results end-to-end with an independent implementation uncovers corruption in the published reference, and discrepancies between our results and the original ones, some of which are not trivial to resolve or remain open questions.
	We further perform a preliminary comparison, not previously explored, of \texttt{ASK}- and
	\texttt{COUNT}-based source selection in the FedX algorithm.
	This work strengthens an already valuable community resource by aligning its artifacts with the RDF standards.
	In doing so, we broaden the set of engines that can be fairly and reproducibly compared.
	We also raise the question of how the results of federated queries under automatic source selection can be made reproducible.
\end{abstract}

\begin{keywords}
    Scientific Reproducibility \sep
    Benchmark \sep
	Linked Data \sep
	RDF \sep
	RDF Repairs \sep
	Interoperability
\end{keywords}

\maketitle

\section{Introduction}
\label{sec:introduction}

Answering difficult questions over the Web of data frequently requires combining facts that reside in separate, independently published datasets.
In the life sciences, for instance, a single question may link a drug in DrugBank to the pathways it participates in from KEGG and the chemical entities that characterize it in ChEBI, while a general-knowledge question may join an entity in DBpedia with its geographic context in GeoNames.
Federated query processing serves precisely this need, by evaluating a single SPARQL query across multiple endpoints at once and integrating their data at query time without centralization~\cite{schwarte2011fedx}.
In practice, replicating every relevant dataset into a single store is often untenable, because the sources sit behind separate administrative boundaries, carry incompatible licenses~\cite{Moreau2021}, and can be too voluminous to copy~\cite{lubiana2025split}.

Reproducibility is a long-standing concern across the sciences~\cite{baker2016reproducibility}, and computer science is no exception~\cite{collberg2016repeatability}.
Benchmarking is an essential part of developing and comparing SPARQL engines and federation algorithms.
Benchmarks are standardized and reproducible scenarios that capture either realistic workloads or interesting choke points in query processing.
For a benchmark to serve as a basis for comparison, independent implementations must be able to ingest both its data and the scenarios it prescribes, a property we call the interoperability of the benchmark.

The quality of published \ac{RDF} data is a studied concern.
The literature contains surveys that organize such concerns into taxonomies of quality dimensions, among them syntactic validity~\cite{Zaveri2015}.
Large-scale empirical studies have in turn measured how far published data deviates from the standards, and cataloged the recurring errors~\cite{Hogan2010WeavingTP,hogan2012empirical}.
Existing tooling addresses such defects at authoring time, where a language server flags syntax mistakes before publication~\cite{Vercruysse2025}, or at ingestion time, where tools such as LOD Laundromat~\cite{Beek2014} republish cleaned copies of existing datasets.
The latter proceeds by exclusion, however, keeping only documents whose source URL parses under RFC~3986~\cite{ietf3986URI}.

This matters especially for a federated benchmark, whose datasets are distributed across multiple endpoints that may each run a different engine.
Such experiments usually compare client-side federation engines, yet the same datasets could equally be used to compare the server-side engines that host them.
In either case, every hosting engine must be able to ingest the benchmark's data, which requires that data to be interoperable.
Interoperable data also lets the benchmark be embedded in automated processes such as \ac{CI} pipelines that track an engine's performance over time.

LargeRDFBench~\cite{saleem2018largerdfbench}, which extends the earlier FedBench~\cite{schmidt2011fedbench} suite, is a widely used benchmark for SPARQL endpoint federation.
More recent benchmarks such as FedShop~\cite{dang2023fedshop} instead generate synthetic sources to stress-test how federation engines scale with the number of endpoints, a concern orthogonal to the standards-conformance of real, published data that we address here.
LargeRDFBench provides 13 real datasets, most of them interlinked, together with 32 queries and their expected results.
Several of these datasets, however, deviate from the \ac{RDF} standard, so an engine that parses \ac{RDF} strictly cannot ingest them.
This is not harmless even for client-side evaluation, since we find the benchmark's expected results to depend on the engine chosen to host the data as well as on the federation semantics the client itself assumes.
It also narrows the set of engines able to host the benchmark and rules out any that parse \ac{RDF} strictly, undermining interoperability and the reproducibility of results using independent implementations.
The expected results compound the problem, since they are published in a custom format that must be parsed with custom tooling, even though standard result formats already exist~\cite{sparql11resultsjson}.
They also contain discrepancies with respect to the source datasets, introduced by the tooling that produced them.

In this paper, we identify and categorize these standards violations and repair them with a reproducible pipeline, producing a standards-conformant edition of LargeRDFBench whose datasets all load under a strict \ac{RDF} parser and whose expected results are re-encoded in a standard format, archived on Zenodo so that it can be downloaded directly~\cite{tam2026lrbedition}.\footnote{\url{https://doi.org/10.5281/zenodo.22140177}}
Running the modernized benchmark end-to-end in a federated engine, we validate the expected results against the source datasets and correct their discrepancies, backing each correction with a reproducible verification script.
This validation also reveals a more subtle threat to reproducibility, as the engine hosting the data can itself change the answers the benchmark returns, independently of the client-side federation algorithm under test.
It further shows that reproducing such results with an independent implementation is difficult, since FedQPL~\cite{Cheng2021}, the reference formalism for automatic source selection, is defined under set semantics and so leaves the real SPARQL execution of automatic source selection ambiguous.
We also explore preliminarily whether the underlying engine running a SPARQL endpoint might have an impact, by comparing \texttt{ASK}- and \texttt{COUNT}-based source selection in the FedX algorithm~\cite{schwarte2011fedx}, as implemented in the Comunica query engine~\cite{taelman_iswc_resources_comunica_2018}, over endpoints hosted by QLever~\cite{bast2017qlever}, a comparison not previously explored.
Because endpoints implement \texttt{ASK} and \texttt{COUNT} differently, this evaluation raises, without settling, the question of how the hosting engine shapes federated performance, a server-side view that current benchmarking overlooks and that the interoperable benchmark now makes possible.

\section{Modernization Process}
\label{sec:method}

We develop a pipeline that transforms the datasets of LargeRDFBench into an interoperable edition.
The pipeline is open source\footnote{\url{https://github.com/shape-federated-queries/large-rdf-bench-modernization}} and fully reproducible, using GNU~make\footnote{\url{https://www.gnu.org/software/make/}} as its orchestrator.
It first downloads the original datasets, then cleans them with software we developed, which also records the repairs it applies as per-dataset statistics in CSV form.
We then use \ac{SOP}~\cite{Champin2026} to serialize every cleaned dataset as N-Triples, and to confirm that each is syntactically valid.
Next, we re-encode the benchmark's expected results in the W3C SPARQL 1.1 Query Results JSON Format with an open-source tool we built for LargeRDFBench\footnote{\url{https://github.com/shape-federated-queries/large_rdf_bench_result_json_format}}, applying to them the same repairs as to the datasets so the two stay aligned.
We also drop the original results' unbound placeholder, an unbound \texttt{OPTIONAL} variable recorded as the literal \texttt{'null'}, which a standards-compliant engine never returns, leaving the variable unbound instead.
We verify that each cleaned dataset preserves the triple count of its original and that each converted result set preserves its number of solutions, emitting a CSV report of this check.
We then convert each dataset into \ac{HDT}~\cite{FERNANDEZ201322}, a compressed RDF format, and run simple \texttt{ASK} queries over it with the Comunica SPARQL engine~\cite{taelman_iswc_resources_comunica_2018}, which parses only valid RDF and so gives an engine-level check that the data loads.
The pipeline thus yields the standards-conformant datasets, their \ac{HDT} counterparts, and the expected results in SPARQL Results JSON.
To make the interoperable edition directly available to the community, we archive it on Zenodo under a persistent identifier~\cite{tam2026lrbedition}: the repaired datasets, their \ac{HDT} indexes, the queries and the expected results can be downloaded and used as they are, without running the pipeline.
We have additionally opened a pull request contributing these artifacts to the official LargeRDFBench repository.\footnote{\url{https://github.com/dice-group/LargeRDFBench/pull/4}}

\subsection{Taxonomy of the Repairs}

We group the standards violations we found into three classes according to the part of the \ac{RDF} syntax they breach, and repair each with a deterministic, reproducible pipeline.\footnote{%
The pipeline additionally applies two repairs for tooling compatibility rather than standards conformance.
It replaces NUL bytes with spaces, which HDT's \texttt{rdf2hdt}~\cite{hdtcpp} cannot ingest, and converts single-quoted RDF/XML attributes to double quotes, which some parsers reject.
Both are omitted from Table~\ref{tab:fixes}.}
We present the three classes in turn, giving for each the repairs it requires and a representative example.
Table~\ref{tab:fixes} reports how often each repair fired across the benchmark's datasets.

\begin{table}[t]
	\centering
	\caption{Standards-conformance repairs applied to LargeRDFBench, aggregated
		over its 13 datasets. Counts are per repair type, so a single
		term may receive several.}
	\label{tab:fixes}
	\begin{tabular}{lrr}
		\toprule
		Repair                                      & Count  & Datasets \\
		\midrule
		\multicolumn{3}{l}{\textit{IRI}}                                \\
		\quad Illegal characters percent-encoded    & 93,587 & 7        \\
		\quad Relative CURIE-like terms absolutised & 1,145  & 1        \\
		\quad Invalid authority colons encoded      & 1,147  & 2        \\
		\quad Leading/trailing spaces stripped      & 76     & 3        \\
		\midrule
		\multicolumn{3}{l}{\textit{Literal}}                            \\
		\quad Multiline literals joined             & 10     & 1        \\
		\quad UTF-16 surrogate pairs combined       & 279    & 1        \\
		\quad Unquoted literals quoted              & 940    & 2        \\
		\midrule
		\multicolumn{3}{l}{\textit{Language}}                           \\
		\quad Malformed language tags fixed         & 9      & 1        \\
		\bottomrule
	\end{tabular}
\end{table}

The first class concerns \acp{IRI} that violate the \ac{IRI} syntax of RFC~3987~\cite{ietf3987IRI}, which builds on the URI generic syntax of RFC~3986~\cite{ietf3986URI}.
Characters that an \ac{IRI} may not contain, such as spaces, control characters and brackets, are percent-encoded; for example, \texttt{ATFIP1[V]} becomes \texttt{ATFIP1\%5BV\%5D}.
A colon in the authority\footnote{The authority as defined by the \emph{Uniform Resource Identifier (URI): Generic Syntax} specification~\cite{ietf3986URI}.} that is not followed by a decimal port is encoded to \texttt{\%3A}, since RFC~3986 permits only digits there.
A term written without a scheme is made absolute against the dataset's base \ac{IRI}; for instance, \texttt{<bio2rdf\_dataset:\ldots>} becomes \texttt{<http://bio2rdf\_dataset\%3A\ldots>}.
Leading and trailing spaces are stripped.

The second class concerns string literals that a strict parser rejects.
A literal broken across physical lines is rejoined with the breaks escaped as \texttt{\textbackslash n}, since a quoted literal may not contain a raw newline in N-Triples~\cite{w3NTriples}; a value spanning three lines, for example, becomes \texttt{"line one\textbackslash nline two\textbackslash nline three"}.
Turtle~\cite{w3Turtle} does admit raw newlines, but only inside a long string delimited by \texttt{"""}, a form N-Triples does not provide, so escaping is required for the serialization we produce.
Both serializations also require every document to be encoded in UTF-8~\cite{w3NTriples,w3Turtle}, so a value that UTF-8 cannot express cannot appear in a conformant file.
A character too large for a single UTF-16 code unit is represented as a surrogate pair, and some literals carry such a character as its two surrogate halves rather than as the character itself.
Because UTF-8 cannot encode surrogates and requires them to be decoded to their character number first~\cite{ietf3629UTF8}, we recombine the halves; the pair \texttt{\textbackslash uD835\textbackslash uDFB1}, for example, becomes the single character U+1D7B1.
A bare alphabetic object, which Turtle reads as the start of a prefixed name~\cite{w3Turtle}, is quoted; for example, \texttt{t:chromosome X} becomes \texttt{t:chromosome "X"}.

The third class concerns language tags that do not conform to BCP~47~\cite{ietf5646BCP47}.
Such a tag is reduced to its primary language subtag; the malformed \texttt{fr\_1793}, for example, becomes \texttt{fr}.

\section{Experiment}
\label{sec:experiment}

The primary purpose of this experiment is to validate the integrity of the modernized benchmark.\footnote{The experiment is reproducible and open source: \url{https://github.com/shape-federated-queries/ask-count-fedx-large-rdf-bench}}
We then illustrate the kind of investigation the now standards-conformant, interoperable benchmark makes possible.
As an example, we compare the \texttt{ASK}- and \texttt{COUNT}-based source-selection strategies of the FedX algorithm~\cite{schwarte2011fedx}.
FedX determines which endpoints can contribute to a triple pattern by sending each of them an \texttt{ASK} query, which an endpoint may answer as soon as it finds one match.
The \texttt{COUNT} variant replaces these \texttt{ASK} queries with \texttt{COUNT} queries, which ask the same question but also return cardinality information that the query planner would otherwise have to estimate.
We report \emph{query planning time} and \emph{query execution time} for both variants.
Throughout, we refer to queries by their LargeRDFBench identifiers~\cite{saleem2018largerdfbench}, where each query can be consulted in full.

We ran the experiment on 14 machines of the imec Virtual Wall testbed, one for each of the 13 endpoints and one client, communicating over the testbed's dedicated internal network, so the times we report are those of a local-area deployment rather than of the public Web.
Every endpoint is served by QLever~\cite{bast2017qlever}, and the client evaluates the queries with the FedX algorithm as implemented in Comunica~\cite{taelman_iswc_resources_comunica_2018}, under both its \texttt{ASK}- and \texttt{COUNT}-based source selection.
Each machine has the same specification: two hexa-core Intel Xeon E5645 (2.4\,GHz) CPUs, 24\,GB of RAM, and Ubuntu 20.04 LTS 64-bit.
We did not run the full query suite, but its simple (S) and complex (C) queries, and allowed each 30 minutes before aborting it.

We corrected a pervasive defect in the original expected results, accented characters recorded as \texttt{?} or the replacement character U+FFFD, verified against the source data.
With this correction applied, every query our engine answers matches the expected results, except three: S7, C8 and C7.
For S7, the triple pattern binding \texttt{"California"} is answered by both NYT and GeoNames; the original results record a single solution, whereas our engine returns two.
FedQPL~\cite{Cheng2021}, the reference formalism for query plans that automatic source selection produces, requires the answer to a query over a federation to be the one obtained by evaluating it over the set union of the data of all federation members~\cite[Def.~3]{Cheng2021}, that is, over the federation seen as a single knowledge graph, in which the triple asserted by both sources occurs once.
Its operators are correspondingly defined over sets of solution mappings~\cite[Def.~6]{Cheng2021}, whereas SPARQL operators evaluate under bag semantics, and the formalism proposes no union operator for SPARQL that would bridge the two, so how to reconcile them is not settled.
A bag semantics for FedQPL is left to future work, and no published extension develops one.
The consequence for reproducibility is significant.
No established formalism describes how federated (real) SPARQL queries under automatic source selection are actually executed, so the answers an engine returns cannot be checked against any definition.
We therefore report the discrepancy rather than resolve it.
For C8, the source data contains the author label both with and without a trailing space (\texttt{"Eyal Oren "} and \texttt{"Eyal Oren"}), so our engine returns both where the original results kept only the trimmed form.
For C7, the difference comes from the hosting engine, not the answer set.
QLever canonicalizes numeric literals, casting \texttt{xsd:float} and \texttt{xsd:double} to \texttt{xsd:decimal}.
SPARQL evaluates over RDF terms, and the dataset asserts the coordinate under both \texttt{xsd:float} and \texttt{xsd:double}, whose use in RDF is itself a documented source of data-quality problems~\cite{keil2022floating}.
The two are therefore distinct solutions, yet QLever collapses them into a single value absent from the data, returning fewer results than the term-based semantics prescribe.
Losing valid solutions this way bears directly on interoperability and reproducibility, as the same benchmark then returns different answers across engines.
A proof file for each case is provided in the analysis repository.~\footnote{\url{https://github.com/shape-federated-queries/large-rdf-bench-result-analysis}}

Comparing the two source-selection strategies, \texttt{ASK} is faster on the median (Table~\ref{tab:qlever-ask-count-adhoc}), but only in aggregate.
The per-query \texttt{ASK}/\texttt{COUNT} execution ratio ranges from 0.15 to 123.62, so on some queries \texttt{ASK} is instead orders of magnitude slower.
This wide spread makes the strategy, and its interaction with the hosting engine, worth investigating, though beyond this paper's scope.

\begin{table}[t]
	\centering
	\caption{Median, minimum and maximum per-query planning and execution time on the
		QLever-hosted federation; the last row is the per-query \texttt{ASK}/\texttt{COUNT}
		ratio (dimensionless).}
	\label{tab:qlever-ask-count-adhoc}
	\begin{tabular}{lrrrrrr}
		\toprule
		                            & \multicolumn{3}{c}{Planning (ms)} & \multicolumn{3}{c}{Execution (s)} \\
		\cmidrule(lr){2-4}\cmidrule(lr){5-7}
		                            & Med & Min & Max & Med & Min & Max \\
		\midrule
		\texttt{ASK}                & 274 & 149 & 665 & 0.6 & 0.3 & 1727.6 \\
		\texttt{COUNT}              & 340 & 226 & 2183 & 1.8 & 0.3 & 32.8 \\
		\texttt{ASK}/\texttt{COUNT} & 0.84 & 0.07 & 1.04 & 0.79 & 0.15 & 123.62 \\
		\bottomrule
	\end{tabular}
\end{table}

\section{Conclusion}
\label{sec:conclusion}

Reproducibility is a cornerstone of science, and of benchmarking in particular, yet a benchmark can uphold it only if every conformant implementation is able to run it.
Several of LargeRDFBench's datasets deviate from the \ac{RDF} standard and its expected results are distributed in an ad hoc format, so we repaired both with a deterministic, reproducible pipeline that yields standards-conformant datasets, their \ac{HDT} counterparts, and the expected results in the W3C SPARQL 1.1 Query Results JSON Format.
We also validated end-to-end in a federated engine, correcting encoding discrepancies in the reference along the way.

Because every conformant engine can now host the data, the repaired benchmark supports reproducible cross-engine comparison.
Reproducing the expected results nonetheless exposed three obstacles to it, of increasing depth.
The first is the ingestion of strings, where, for example, trailing spaces kept by one tool and trimmed by another change which literals exist.
This is trivial to resolve once noticed.
The second is that an engine may materialize entailments as it ingests the data.
Collapsing \texttt{xsd:float} and \texttt{xsd:double} into \texttt{xsd:decimal}, for example, can change the query results, including the multiplicity of the bag of solution mappings.
The third is the absence of a fundamental definition, since for federated queries under automatic source selection no formalism describes what a SPARQL engine should compute.
FedQPL~\cite{Cheng2021}, which formalizes the plans that automatic source selection produces, defines their execution over sets of solution mappings, whereas SPARQL evaluates under bag semantics.
Extending FedQPL, or establishing another formal definition of federated queries with automatic source selection under bag semantics, is in our view an important open problem for reproducible federated evaluation.
Our comparison of \texttt{ASK}- and \texttt{COUNT}-based source selection opens a further direction, as the strategy's cost varies significantly across queries and depends on how the hosting engine evaluates \texttt{ASK} and \texttt{COUNT}, a real-world constraint on federation that current benchmarking practice overlooks.
We leave a systematic multi-engine evaluation to future work.

\begin{acknowledgments}
	This research was supported by Serendipity Engine (Research Foundation - Flanders (FWO) grant number S006323N).
	Ruben Taelman is a postdoctoral fellow of the Research Foundation – Flanders (FWO) (1202124N).
\end{acknowledgments}

\section*{Declaration on Generative AI}
During the preparation of this work, the authors used LLM technologies in order to:
Drafting content and Code generation.
After using these tools/services, the authors reviewed and edited the content
as needed and take full responsibility for the publication's content.

\bibliography{references}

\end{document}